\documentclass[11pt]{article}
\usepackage{hyperref}
\pdfoutput=1
\begin{document}
\title{Dynamics of Surfactants Spreading on Gel-like Materials: \\ Cracking and Pattern Formation}
\author{Constantine Spandagos, Paul F. Luckham and Omar K. Matar \\
\\\vspace{6pt} Department of Chemical Engineering, \\ Imperial College London, London, SW7 2AZ, UK}
\maketitle
\begin{abstract}
\begin{center}
Fluid Dynamics Video \\
\end{center}

\vspace{2pt}\noindent We study the dynamics of surfactants spreading
on gels, paying particular attention to the pattern formation
accompanying the flow. The latter results from gel-cracking,
promoted by Marangoni stresses, and resemble ``starbursts".
\end{abstract}
\section{Introduction}
The deposition and spreading of drops of surfactant solutions on the
surface of gels gives rise to the shaping of crack-like spreading
``arms" in formations that resemble ``starbursts". Marangoni
stresses induced by surface tension gradients between the spreading
surfactant and the uncontaminated gel layer are identified to be the
main driving force behind the observed phenomena. Examples of these
formations are shown in
\href{http://workspace.imperial.ac.uk/ceFluidMechanics/Public/spandagos.mov}{Video1}.

The morphology and spatio-temporal evolution of these "starburst"
patterns is examined. We use drops laden with two types of
surfactants: SDS (sodium dodecyl sulphate), and Silwet L-77
(polyalkyleneoxide modified heptamethyltrisiloxane- a superspreader)
in concentrations ranging from 0.3cmc to 100cmc. The underlying gel
substrates are made of agar (a polysaccharide-based gel), and
gelatine (a protein-based gels). Agar is used in concentrations from
0.04 wt\% to 0.14 wt\%, and gelatine is used in concentrations from
0.7 wt\% to 0.17 wt\%.

%
\end{document}